\documentclass[prd,amsmath,amsfonts,showpacs,preprintnumbers,%
               superscriptaddress,floatfix,%
               twocolumn,a4paper]{revtex4}

\usepackage{graphicx}

\newcommand{\Eq}[1]{Eq.~(\ref{#1})}
\newcommand{\Fig}[1]{Fig.~\ref{#1}}

\renewcommand{\Re}{\operatorname{\mathfrak{Re}}}    
\newcommand{\Tr}{\operatorname{Tr}}                 
\newcommand{\tr}{\operatorname{tr}}                 
\newcommand{\cSB}{D$\chi$SB}
\newcommand{\DcSB}{D$\chi$SB}

\usepackage{color}
\newcommand{\hbo}{\hbox to 1cm {\hfill }}

\begin{document}

\preprint{ADP-08-04/T664}

\title{Center vortices and the quark propagator in SU(2) gauge theory}

\author{Patrick~O.~Bowman}
\affiliation{Centre of Theoretical Chemistry and Physics, 
  Institute of Fundamental Sciences, Massey University (Auckland), 
  Private Bag 102904, NSMSC, Auckland NZ}
\author{Kurt~Langfeld}
\affiliation{School of Maths \& Stats, University of Plymouth,
  Plymouth, PL4 8AA, England}
\author{Derek~B.~Leinweber}
\author{Alan~O'~Cais}
\author{Andr\'e~Sternbeck}
\author{Lorenz~von~Smekal}
\author{Anthony~G.~Williams} 
\affiliation{Centre for the Subatomic Structure of Matter 
  (CSSM), School of Chemistry \& Physics, University of Adelaide 5005,
  Australia} 

\date{September 9, 2008}

\begin{abstract}
  We study the behavior of the AsqTad quark propagator in Landau gauge
  on quenched SU(2) gauge configurations under the removal of center
  vortices. In contrast to recent results in SU(3), we clearly see the
  infrared enhancement of the mass function disappear if center
  vortices are removed, a sign of the intimate relation between center
  vortices and chiral symmetry breaking in SU(2) gauge-field theory.
  These results provide a benchmark with which to interpret the SU(3)
  results. In addition, we consider vortex-only configurations. On
  those, the quark dressing function behaves roughly as on the full
  configurations, and the mass function picks up an almost linear
  momentum dependence.
\end{abstract}

\maketitle

\section{Introduction}

Dynamical breaking of chiral symmetry (\cSB) is an essential
nonperturbative property of quantum chromodynamics (QCD) which cannot
be accounted for within perturbation theory at any order. Only
nonperturbative approaches, such, as those provided by lattice QCD
simulations or studies of the Dyson-Schwinger equations, can be used
to explore this phenomenon.

The other characteristic nonperturbative phenomenon of QCD is
confinement: the fact that colored states are never observed.  It is
tempting to speculate that these two phenomena might be driven by the
same basic mechanism, an idea supported by finite-temperature studies
where the deconfinement and chiral restoration transitions are
observed to occur at coincident
temperatures~\cite{Laermann:2003cv}. Moreover, it was found that the
low-lying modes of the quark operator not only bear witness on
spontaneous chiral symmetry breaking but also on
confinement~\cite{Synatschke:2008yt}.

One leading candidate for such a mechanism is the center vortex.
Center vortices have been studied in lattice QCD simulations for more
than a decade.  The recovery of the string tension from
``vortex-only'' SU(2) gauge configurations [i.e., $\text{Z}_2$
projected from SU(2)] is well known, as is the recovery of the chiral
condensate~\cite{deForcrand:1999ms,Alexandrou:1999iy,Langfeld:2003ev,
Gattnar:2004gx} 
(see also more recent
Refs.~\cite{Boyko:2006ic,Bornyakov:2007fz,Hollwieser:2008tq}).  In
SU(3), however, the situation is far less compelling.  For example, in
Refs.~\cite{Leinweber:2006zq,Bowman:2008tobepublished} numerical
evidence has been given that in SU(3) mass generation remains intact
after removing center vortices, whereas the string tension vanishes as
expected.  This came as a surprise and immediately suggested a
corresponding ``bench-mark'' study for the case of SU(2) which is
reported here.

We use the quark propagator as a probe of \DcSB{}.  The Dirac scalar
part of the propagator, related at large momenta to the perturbative
running mass, is enhanced at low momenta, even in the chiral
limit~\cite{Bowman:2005zi}: a demonstration of \DcSB{}.  We explicitly
establish the relation between center vortices and \cSB{} by
investigating the quark propagator under the removal of center
vortices. Specifically, we will provide numerical evidence that
dynamical mass generation disappears if those vortices are removed,
and surprisingly, much of it resides in the vortex-only
part. Additionally, we present an improved method for generating the
SU(2) maximal-center-gauge-projected configurations.

\medskip

The paper is organized as follows. In an attempt to make it
self-contained we briefly introduce the quark propagator and its
realization on the lattices based on the AsqTad quark action. This is
followed by a specification on how we gauge-fixed and
identified center vortices in our gauge configurations. Then, results
for the mass and quark dressing function are compared on full,
vortex-removed and vortex-only configurations. A summary concludes the
paper.

\section{The quark propagator}

The quark propagator is gauge dependent. In covariant gauges in
Euclidean momentum space it can be parametrized in the general form
\begin{equation} S(p^2) = \frac{Z(p^2)}{i\gamma\cdot p + M(p^2)}
\end{equation} where $M$ is the running mass and $Z$ the quark
dressing function. $S$ can be calculated in regularized theories, as
for example here in a lattice regularization where the lattice spacing
$a$ makes all expressions finite. At sufficiently small $a$, i.e., if
scaling violations due to finite lattice spacings are negligible, the
bare quark propagator, $S$, is related to the renormalized propagator
via multiplicative renormalization:
\begin{displaymath} S_R\big(p^2; \mu, g_R(\mu), m_R(\mu)\big) = Z_2\,
S\big(p^2; a, g_0(a),m_0(a)\big).
\end{displaymath} To ensure multiplicative renormalizability, all the
dependence of $S_R$ on the renormalization point $\mu$ is contained in
the renormalized quark dressing function, $Z_R$. $M$ does not depend
on $\mu$. A renormalization condition fixes $Z_2$, the renormalization
constant of the quark fields. Lattice calculations often use MOM
schemes to fix renormalization constants. In MOM schemes, $Z_2$ is
fixed by requiring $S_R$ to be of the form of a free propagator at the
renormalization point $p^2=\mu^2$. This sets
\begin{equation} Z_R(\mu^2,\mu^2) = 1 \quad\text{and}\quad M(\mu^2) =
m_R(\mu)
\end{equation} where the latter denotes the renormalized mass at
$\mu^2$.

Calculation of the quark propagator $S$ proceeds like any correlation
function in a lattice Monte Carlo (MC) calculation once the gauge has
been fixed.  For the gauge we used the ever popular Landau gauge. It
is straightforward to implement on the lattice and allows for an easy
comparison to other studies.  The Landau-gauge quark propagator has
been studied widely in SU(3) gauge theory using Wilson-clover,
staggered-type and overlap actions in quenched and unquenched
simulations (see, e.g, Refs.~\cite{Skullerud:2000un,Skullerud:2001aw,
Bowman:2002bm,Bowman:2002kn,Bonnet:2002ih,Zhang:2003faa,Parappilly:2005ei,
Bowman:2005zi,Kamleh:2007ud}). It has been shown that the quark
propagator obtained with the AsqTad-improved staggered quark action
possesses good symmetry properties and is well behaved at large
momenta (see, e.g., Ref.~\cite{Bowman:2004xi}).  The \mbox{AsqTad} action was
therefore a natural choice for this study.

\section{Details of the calculation}

Configurations were generated on a $16^3\times32$ lattice using a
tadpole-improved Wilson gauge action with an inverse coupling constant 
$\beta=1.35$.  Around 120 configurations were used.

\subsection{Identifying center vortices}

\begin{figure}
\includegraphics[width=0.9\linewidth]{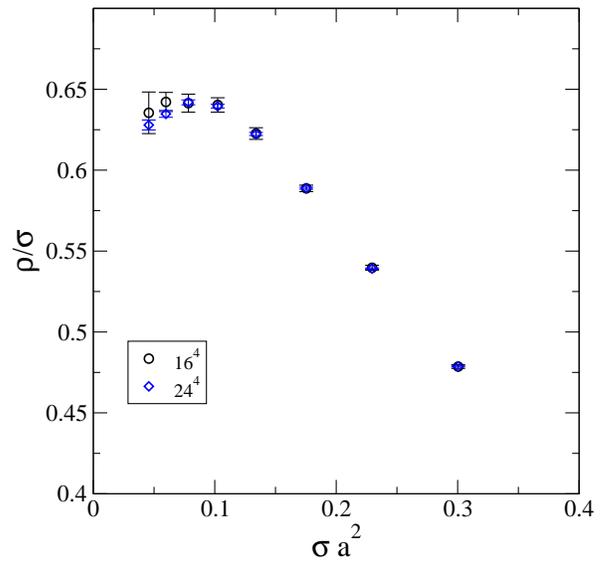} 
\caption{Scaling of the planar vortex area density $\rho$  in units
  of the string tension $\sigma $ as function of the lattice spacing
  $a$.}
\label{fig:density2} 
\end{figure}
The observation that the long-range static potential only depends on
the center charge (also called ``$N$-ality'') of the quark
representation led to the expectation that the center subgroup $Z_N$
of SU($N$) plays a crucial role for quark confinement.  Center
vortices emerge from the corresponding $Z_N$ gauge theory: they form
closed worldsheets in four space-time dimensions carrying flux which
takes values in the center of the SU($N$) group.  Early attempts to
define this subgroup by the projection SU($N$) $\rightarrow Z_N$ failed
in the sense that the arising vortex matter did not have meaningful
properties in the continuum limit of vanishing lattice spacing $a$. In
the pioneering works~\cite{DelDebbio:1996mh,DelDebbio:1998uu} a two
step process was proposed to define the links $Z_\mu (x)$ spanning the
$Z_N$ gauge theory:
\begin{eqnarray}
  &\hbox{(i)}& \quad 
  \sum_{x, \mu } \vert \Tr U^g_{\mu} (x) \vert^2 \, 
  \stackrel{g}{\longrightarrow }\,  \max 
  \label{eq:k1} \\
  &\hbox{(ii)}& \quad \Re\Tr \Bigl( U^g_\mu (x) \, Z^{\dagger}_{\mu} 
  (x) \Bigr) \stackrel{g}{\longrightarrow }\,  \max \; . 
  \label{eq:k2} 
\end{eqnarray}
Step {(i)} is the difficult part involving gauge fixing and the
interference of Gribov ambiguities (see comments below). The
projection step {(ii)} operates locally and can be implemented
straightforwardly. Using a standard iteration overrelaxation
procedure~\cite{DelDebbio:1998uu} for step {(i)} finally defines
vortex matter with a sensible phenomenology in the continuum limit:
for SU(2), it was observed that the so-called ``vortex-only''
configurations, defined by the center links $Z_N(x)$ reproduce a good
deal of the string tension while ``vortex-removed'' configurations,
spanned by the links
\begin{equation}
  \widetilde{U}_{\mu}(x)  = Z^{\dagger}_{\mu} (x) \; U_{\mu}(x) \; , 
  \label{eq:k3}
\end{equation} 
do not support a linear rising static potential at large
distances~\cite{DelDebbio:1996mh,DelDebbio:1998uu,Greensite:2003bk}.
It was put forward in~\cite{Langfeld:1997jx} that this heuristically
defined vortex texture has sensible properties in the continuum limit.
It was subsequently discovered that these vortices provide an
intriguing picture of the deconfinement phase transition at finite
temperatures~\cite{Engelhardt:1999fd}. They even admit detailed
insights in the critical
phenomenon~\cite{Langfeld:2003zi,Langfeld:2005kp} in complete
agreement with studies of the free energy of gauge-invariant center
vortices over this transition \cite{deForcrand:2001nd}.

We point out that present day algorithms~\cite{Giusti:2001xf} are only
capable of obtaining one of the many local maxima of the gauge-fixing
condition (\ref{eq:k1}) and that the vortex properties do depend on
the set of maxima~\cite{Kovacs:1999st,Bornyakov:2000ig,Faber:2001hq}.
Localizing the global maximum of \Eq{eq:k1} would remove this
ambiguity, but it is not clear whether the vortex matter arising from
the global maximum would be of any phenomenological value.  In fact,
an anticorrelation was found~\cite{Bornyakov:2000ig}: the larger the
gauge-fixing functional (\ref{eq:k1}), the lesser the string tension
obtained from vortex-only configurations.  Although we have not
yet found a concise mathematical description, the vortex matter best
for phenomenological studies seems to arise from an average over
Gribov copies.

Note also that the relation between vortices and quark confinement is
far less striking for the gauge group SU(3): despite many attempts,
the string tension arising from vortex-only configurations is
systematically smaller than the full string
tension~\cite{Langfeld:2003ev,Cais:2007bm,Cais:2008}. The difference
in quality between the SU(2) and SU(3) vortex picture indicates that
we are still missing a point for SU(3) and partially motivated the
present investigation of the SU(2) quark propagator.

Because of the above ambiguity, we are going to present 
the details of the gauge-fixing procedure which gives rise 
to an intriguing vortex phenomenology. 
For SU(2), the gauge-fixing matrix $g(x) \in\mathrm{SU(2)}$ can be viewed 
as a 4-dimensional unit-vector in Euclidean color space: 
\begin{displaymath}
  g(x) = g_0 (x) + i \vec{\tau} \vec{g}(x) \; , \qquad
  G(x)  = \left( \begin{array}{c} 
      g_0 (x) \\ \vec{g} (x) \end{array} \right) \; ,  
\end{displaymath}
where
\begin{displaymath}
  g^2_0 (x) + \vec{g}^2(x) = 1\;.
\end{displaymath}
Maximizing the gauge-fixing
functional (\ref{eq:k1}) is an iterative procedure: choosing a
particular site $x_0$ and setting $g(x\not=x_0)=1$, the functional is
locally maximized, and the links are updated accordingly: $U_\mu
\rightarrow U^g_{\mu} $. Subsequently, all sites are visited and many
sweeps through the lattice are performed until the gauge-fixing action
does not change anymore within the required precision. 

For the local update $g(x_0)$, let $s_\mathrm{fix}$ denote that 
part of the gauge-fixing functional (\ref{eq:k1}) which is 
affected by a change of $g(x_0)$:
\begin{displaymath}
  s_{\mathrm{fix}} \; = \; G^T (x_0) \; M \; G(x_0)  \; - \; 
  \lambda \; \Bigl( G^T(x_0) G(x_0) -1 \Bigr) \; ,  
\end{displaymath}
where $\lambda $ is a Lagrange multiplier and $M$ is a real symmetric
$4\times 4$ matrix given in terms of the link fields [$U_\mu (x) \; =
\; u^0 _\mu (x)\; + \; i \vec{\tau} \vec{u}_\mu (x)$]:  
\begin{widetext}
\begin{equation}
M(x) \, = \, \sum _{\mu=1}^4 \, \left( \begin{array}{cc} 
(u^0_\mu (x))^2 + (u^0_\mu (x-\mu))^2 & 
- u^0_\mu (x) u^i_\mu (x) + u^0_\mu (x-\mu) u^i_\mu (x-\mu) \\ 
- u^0_\mu (x) u^i_\mu (x) + u^0_\mu (x-\mu) u^i_\mu (x-\mu) & 
u^i_\mu (x) u^k_\mu (x) + u^i_\mu (x-\mu) u^k_\mu (x-\mu) 
\end{array} \right) . 
\end{equation} 
\end{widetext}
We also introduce the eigenvectors and eigenvalues 
of the matrix $M$:
\begin{displaymath}
  e_k,\;\lambda_k\qquad\text{with}\quad k=1 \ldots 4\;.
\end{displaymath}
Choosing the largest eigenvector for the gauge transformation, i.e., 
\mbox{$G(x_0) = e_{\max}$}, the local increase of the gauge-fixing 
functional is maximized: 
\begin{displaymath} 
  s_{\mathrm{fix}}  = \lambda_{\max} \; . 
\end{displaymath}
This choice gives rise to the standard iteration procedure 
which is usually employed for MCG fixing~\cite{DelDebbio:1998uu}. 

\medskip
Here, we depart from this standard procedure and introduce an aspect
of simulated annealing: we choose $G(x_0) = e_k$ with a relative
probability of $\exp\{\beta_f \lambda_k\}$, where $\beta_f$ is an
auxiliary parameter familiar from simulated annealing. For large
$\beta_f$, the probability for picking the largest eigenvalue is high,
and our method smoothly merges with the standard scheme. In practice,
we started with $\beta_f = 0.02$ and performed 25 sweeps through the
lattice until we increased $\beta_f$ by 0.1. The procedure stopped
when no further increase of the gauge-fixing functional was achieved.

It turns out that the vortex matter arising from this procedure has
good phenomenological properties such as good scaling properties in
the continuum limit. To test the latter aspect, we calculated the
planar vortex area density $\rho$ in units of the (measured) string
tension for several values of the lattice spacing $a$ using the
standard Wilson action. Figure \ref{fig:density2} illustrates how the vortex
density becomes independent of the lattice regulator for sufficiently
small values of the lattice spacing.

\subsection{Fixing to Landau gauge}

To fix configurations to Landau gauge an overrelaxation algorithm was
used. This is an iterative algorithm that maximizes the Landau-gauge
functional 
\begin{equation}
  F_U[g] = \frac{1}{4VN_c}\sum_{x,\mu} \Re\Tr\left[
  g_xU_{x,\mu}g^{\dagger}_{x+\hat{\mu}}\right] 
\label{eg:Landau_gauge_functional}
\end{equation}
by changing the gauge transformation fields $g_x$ locally but keeping
$U$ fixed. The algorithm stopped if an accuracy of
\begin{equation}
  \max_x\Tr\left[\partial_{\mu} A_{x,\mu}\partial_{\mu}
  A^{\dagger}_{x,\mu} \right] < 10^{-13}
\label{eq:stopping_criterion}
\end{equation}
was reached 
\endnote{Here a note is in order. Maximizing $F_U[g]$ is not
  unique as there are different $g$'s all satisfying
  relation~(\ref{eq:stopping_criterion}). This ambiguity, known as the
  Gribov-copy problem, has been shown to systematically affect data,
  e.g., of the Landau-gauge ghost propagator, while for others, e.g.,
  for the gluon propagator in the same gauge, the impact stays within
  statistical errors (see, e.g., Ref.~\cite{Sternbeck:2005tk}). 
  Neither does maximization sample all Gribov copies.  In that,
  lattice Landau gauge crucially differs from the continuum Landau
  gauge which is based on a BRST average over {\em all} copies.  The
  modified lattice Landau gauge of \cite{vonSmekal:2007ns}, however,
  provides a framework for lattice BRST without Neuberger $0/0$
  problem which prevented that for 20 odd years.
}.
Here the gauge potential is defined as
\begin{displaymath}
  A_{x,\mu} := \frac{1}{2iag_0}\left(U_{x,\mu} -
  U^{\dagger}_{x,\mu}\right)\;.
\end{displaymath}
For the vortex-only configurations, Fourier acceleration was also used
\cite{Davies:1987vs}. We applied the standard technique where each
configuration was gauge-fixed once.

\subsection{The AsqTad quark propagator}

On those gauge-fixed configurations we calculate the quark propagator
$S(x,y)$ in coordinate space by inverting the AsqTad fermion
matrix. After Fourier-transforming $S(x,0)$, the mass and dressing
functions, $M$ and $Z$, are extracted from $S$ by suitable projections
in Dirac space (see below). At this step it is important to know how
the discrete lattice momenta 
\begin{displaymath}
  p_{\mu}=\pi k_{\mu}/L_{\mu}\qquad\text{with}\quad 
  k_{\mu}\in(-L_{\mu}/2,L_{\mu}/2]
\end{displaymath}
are related to the physical momenta $a^2q^2(p)$ in lattice units. From
experience we know it is always good practice to look at the
tree-level form of the lattice propagators, and to define momenta such
that the continuum tree-level expression is retrieved. This, known as
tree-level correction \cite{Leinweber:1998uu}, accounts for the lowest
order discretization effects.  With the AsqTad action the tree-level
form of the quark propagator in Landau gauge reads
\begin{displaymath}
  S_L^{-1}(a^2p^2) = i\sum_{\mu=1}^4\bar{\gamma}_{\mu} aq_{\mu}(p_{\mu}) + ma
\end{displaymath}
where $p_{\mu}$ is as above and 
\begin{equation}
  aq_{\mu}(p_{\mu}) =
  \sin(p_{\mu})\left[1+\frac{1}{6}\sin^2(p_{\mu})\right]
  \label{eq:definition_of_momentum}
\end{equation}
is the ``kinematic momentum'' (see, e.g., Ref.~\cite{Bowman:2002bm}). The
four matrices $\bar{\gamma}_{\mu}$ form a staggered Dirac algebra
(Eqs.~(A.6) and (A.7) of Ref.~\cite{Bowman:2005vx}). Consequently, we
use \Eq{eq:definition_of_momentum} to define momenta.

The functions $M$ and $Z$ are then extracted from $S_L(p)$ by taking
the two traces in Dirac space
\begin{align}
  A_L(q,a) &= \frac{i}{4N_cq^2a^2}\tr(\bar{\gamma}_{\mu} q_{\mu}\,
 S_L(q,a)) \\
  B_L(q,a) &= \frac{1}{4N_c}\tr S_L(q,a)
\end{align}
and combining them to (here $N_c=2$)
\begin{align}
  Z_L(q,a) &= \frac{q^2a^2 A_L^2(q,a) + B_L^2(q)}{A_L(q,a)}\\
\intertext{and}
  M_L(q,a) &= \frac{B_L(q,a)}{A_L(q,a)}\;.
\end{align}

In the region of asymptotic scaling, $M_L(q,a)$ becomes independent of
$a$ and equals the running mass. $Z$ needs to be renormalized as
explained above. As we are only interested in qualitative changes
of the momentum-dependence of $M$ and $Z$ under the removal of center
vortices, we prefer to work directly with unrenormalized
quantities. Therefore, the presentation of the data is simplified by
considering only bare lattice functions at a fixed lattice spacing.
Consequently, lattice artefacts could not be directly assessed;
  however, for this value of the coupling, discretization errors are
  not expected to be significant.

A cylinder cut \cite{Leinweber:1998uu} is applied to all the data to
reduce the effects of rotational symmetry violation.

\begin{figure*}
  \centering
  \includegraphics[width=\linewidth]{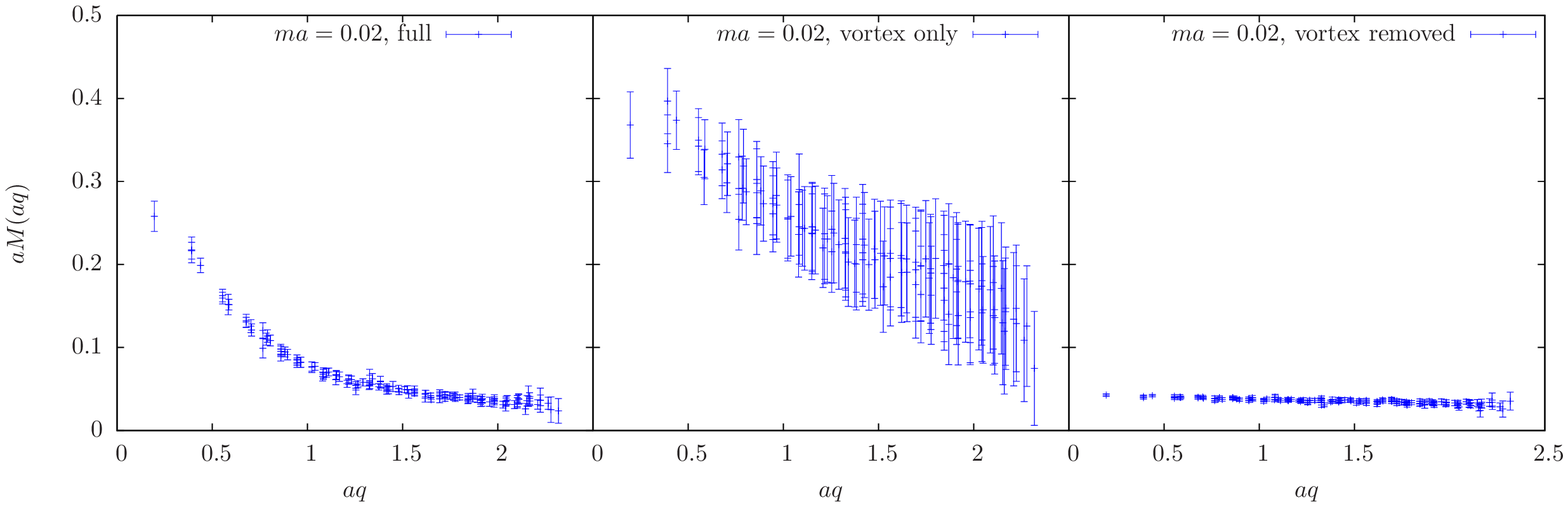}
  \includegraphics[width=\linewidth]{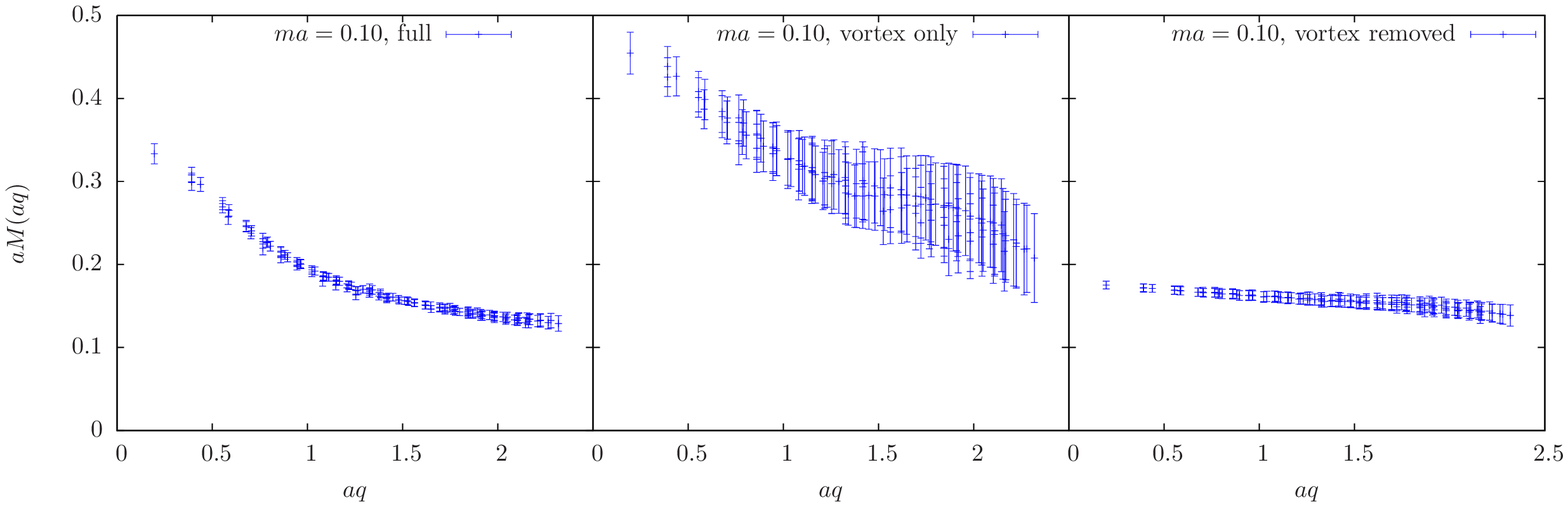}
  \caption{The mass function $aM(q)$ versus kinematic momentum
    for two bare quark masses $ma=0.02$ (top) and $ma=0.10$ (bottom).  We
    show data on full, vortex-only and vortex-removed configurations from
    left to right.  Data has been cylinder cut.  The nonperturbative
    enhancement of the mass function at low momenta is associated with the
    presence of center vortices.}
  \label{fig:Mall}
\end{figure*}

\begin{figure*}
  \includegraphics[width=\linewidth]{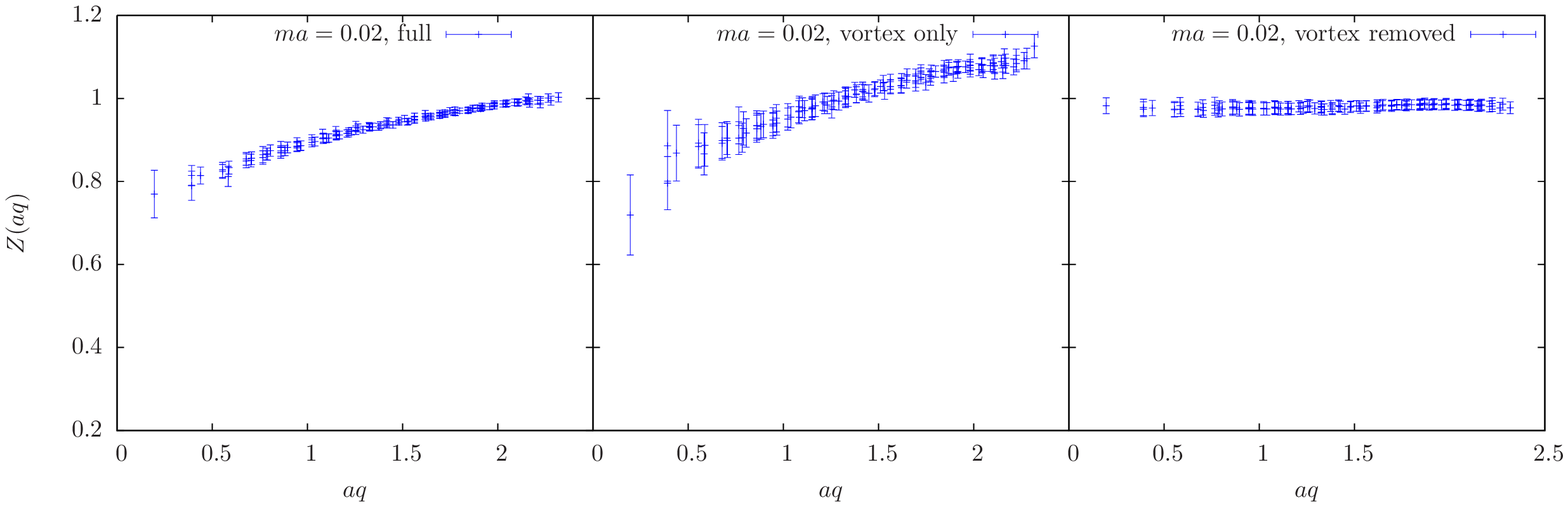}
  \includegraphics[width=\linewidth]{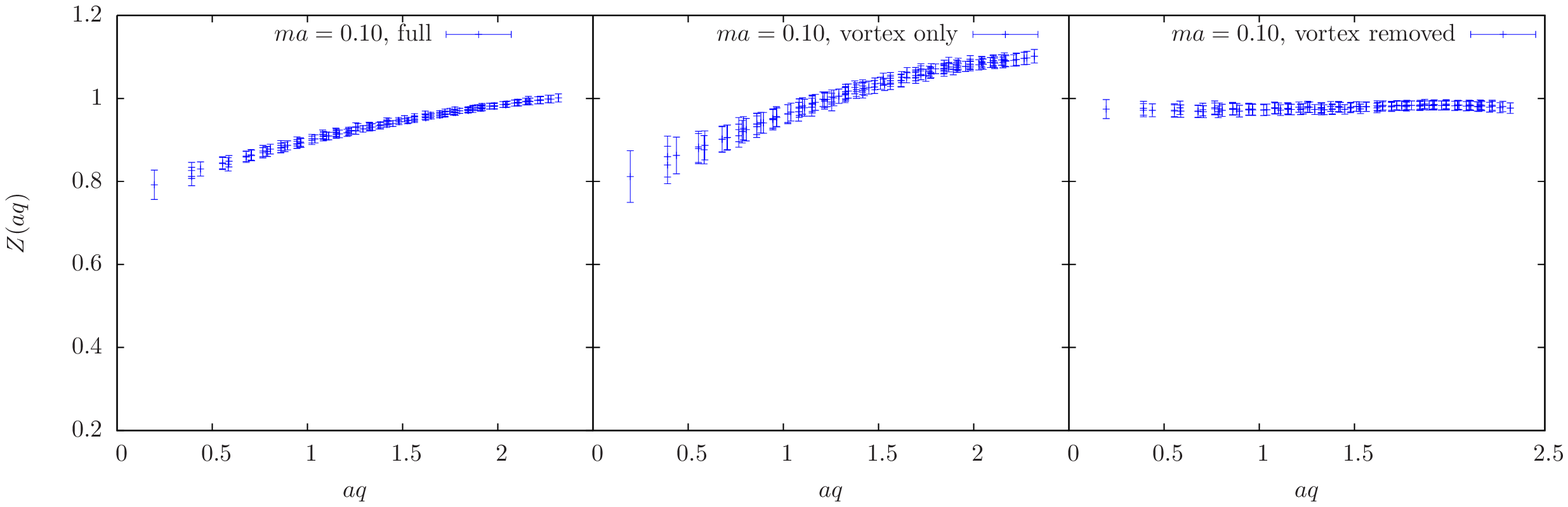}
  \caption{The bare quark dressing function, $Z$, as a function of
    momentum for two bare quark masses $ma=0.02$ (top) and $ma=0.10$
    (bottom).  The left column displays data on full configurations,
    whereas in the middle and right column the data on vortex-only and
    vortex-removed configurations are shown, respectively.  Data has been
    cylinder cut.  The nonperturbative suppression at low momenta is
    associated with the presence of center vortices.}
  \label{fig:Zall}
\end{figure*}

\section{Results}

\subsection{The mass function}

In \Fig{fig:Mall} we compare the mass function on our sets of full,
vortex-only and vortex-removed configurations. Data was obtained for
a range of bare quark masses from $ma=0.020$ to $ma=0.100$, and the
two extremes are shown here as functions of momentum.

Our results on the full SU(2) configurations (see \Fig{fig:Mall}, left
column) show a large enhancement near zero momentum, while data drops
rapidly to its expected asymptotic behavior at large momentum.  Also,
as expected, the infrared enhancement is stronger for the smaller bare
quark mass. That is, our data for $M(q^2)$ on the full configurations
clearly reproduce the well-known characteristics expected for the mass
function in both the nonperturbative and perturbative regime.

On the vortex-removed configurations (see \Fig{fig:Mall}, right
column), the mass function is more or less flat taking values slightly
above $ma$. The momentum dependence of $M$ is almost linear with a 
bigger slope for $ma=0.1$. Thus, the dynamical contribution to the
mass function, which we clearly see on the full configurations,
disappears when center vortices are removed.

Interestingly, the mass function on the vortex-only configurations
(see \Fig{fig:Mall}, middle column) depends quite strongly on
momentum.  Even though the signal is quite noisy, $M(q^2)$ grows
almost linearly upon decreasing $a^2q^2$. That is, much of the
infrared enhancement of $M$ is contained in the vortex-only part which
clearly underlines the importance of center vortices as IR degrees of
freedom.

\subsection{The quark dressing function}

The same comparison for the bare quark dressing function, $Z$, is
shown in \Fig{fig:Zall}. Again, we show data at $ma=0.02$ and
$ma=0.10$ on full SU(2) configuration in the left column, while the
middle and right column displays data on vortex-only and
vortex-removed configurations, respectively. As expected, on the full
configurations, the quark dressing function takes values around one at
large momenta and becomes suppressed towards lower momenta. The
smaller the quark mass the more pronounced the dip at lower
momenta. If center vortices are removed, the infrared suppression
disappears and $Z$ is a flat function of momentum (see the right
column). There, $Z$ roughly stays at its tree-level
value. Surprisingly, on the vortex-only configurations, $Z$ has a
similar momentum dependence to the full configurations.  Again, the
results are noisier, but the infrared suppression is unambiguous.

\section{Conclusions}

We have studied the Landau-gauge quark propagator in quenched SU(2)
gauge theory under the removal of center vortices. Our implementation
of this propagator is based on the AsqTad-improved staggered quark
action modified to SU(2). The full propagator is found to strongly
resemble that of the SU(3) theory.

Our results for the mass and quark dressing functions unambiguously
show the disappearance of \DcSB{} when center vortices are removed.
This is in contrast to the situation in
SU(3)~\cite{Leinweber:2006zq,Bowman:2008tobepublished}. 
There, even after center-vortex removal dynamical mass generation
survives while the string tension is flat.

Additionally, we have studied the quark propagator on vortex-only
configurations. Even though the signal is quite noisy, both parts of
the propagator reveal a form qualitatively similar to the full,
untouched configurations. 

In summary, our SU(2) results clearly represent a strong relationship
between the vortex picture and spontaneous breaking of chiral
symmetry. We do not expect this to be merely a feature of
staggered fermions, however it might be valuable to repeat this study
with another type of fermion, in particular after the new results
presented in Ref.~\cite{Hollwieser:2008tq}. More importantly,
whether it is possible to observe similar behavior in SU(3) [or
ultimately SU($N$)] is still an open question.\vfill

\section*{ACKNOWLEDGMENTS}

POB thanks the CSSM for its hospitality during part of this work. This
research made use of the publicly available MILC code and was
supported by the Australian Research Council, eResearch South Australia
and the Australian Partnership for Advanced Computing.\par\vfill


\end{document}